\documentclass[aps,noshowpacs,amsmath,amssymb,twocolumn,aps,nofootinbib,prx]{revtex4}
\usepackage{graphicx}
\usepackage{natbib}

\DeclareMathOperator{\sech}{sech}
\newcommand{\be}{\begin{equation}}
\newcommand{\ee}{\end{equation}}
\newcommand{\bea}{\begin{eqnarray}}
\newcommand{\eea}{\end{eqnarray}}

\newcommand{\expected}[1]{\left\langle#1\right\rangle}
\newcommand{\var}[1]{\expected{\Delta^2#1}}

\newcommand{\forget}[1]{}

\begin{document}
\title{Transmission Estimation at the Fundamental Quantum Cram\'er-Rao Bound with Macroscopic Quantum Light}
\author{Timothy S. Woodworth$^{1,2}$, Carla Hermann-Avigliano$^{3,4}$, Kam Wai Clifford Chan$^{5}$, and Alberto M. Marino$^{1,2}$}
\affiliation{$^{1}$Homer L. Dodge Department of Physics and Astronomy, The University of Oklahoma, Norman, Oklahoma, 73019, USA}
\affiliation{$^{2}$Center for Quantum Research and Technology, The University of Oklahoma, Norman, Oklahoma, 73019, USA}
\affiliation{$^{3}$Departamento de Física, Facultad de Ciencias Físicas y Matemáticas, Universidad de Chile, Santiago, Chile}
\affiliation{$^{4}$ANID - Millennium Science Initiative Program - Millennium Institute for Research in Optics (MIRO), Chile}
\affiliation{$^{5}$OAM Photonics LLC, San Diego, CA 92126, USA}

\begin{abstract}
The field of quantum metrology seeks to apply quantum techniques and/or resources to classical sensing approaches with the goal of enhancing the precision in the estimation of a parameter beyond what can be achieved with classical resources. Theoretically, the fundamental minimum uncertainty in the estimation of a parameter for a given probing state is bounded by the quantum Cram\'er-Rao bound. From a practical perspective, it is necessary to find physical measurements that can saturate this fundamental limit and to show experimentally that it is possible to perform measurements with the required precision to do so. Here we perform experiments that saturate the quantum Cram\'er-Rao bound for transmission estimation over a wide range of transmissions when probing the system under study with a bright two-mode squeezed state. To properly take into account the imperfections in the generation of the quantum state, we extend our previous theoretical results to incorporate the measured properties of the generated quantum state. For our largest transmission level of  84\%, we show a 62\% reduction over the optimal classical protocol in the variance in transmission estimation when probing with a bright two-mode squeezed state with 8~dB of intensity-difference squeezing. Given that transmission estimation is an integral part of many sensing protocols, such as plasmonic sensing, spectroscopy, calibration of the quantum efficiency of detectors, etc., the results presented promise to have a significant impact on a number of applications in various fields of research.
\end{abstract}
\maketitle

\section{Introduction}

The second quantum revolution seeks to develop new technology that can take advantage of quantum resources and that can lead to practical applications of quantum mechanics. These include, for example,  quantum computing~\cite{Nielsen2000} to solve problems intractable for classical computers, such as factorization~\cite{Shor1997,FernandezCarames2020} and database searching~\cite{Grover1996}; quantum cryptography for transfer of information with absolute security~\cite{Bedington2017,Kumar2021,Pan2021}; quantum imaging for enhanced resolution~\cite{Tsang2016, Kumar2021a, Sorelli2021} for applications such as imaging of biological samples without damage~\cite{Taylor2013,Taylor2014}; and quantum metrology~\cite{Degen2017,Tan2019} for enhanced measurements. In particular, quantum metrology, which is the focus of this paper, seeks to use quantum resources, such as quantum states of light, to enhance systems and measurements beyond what is possible with classical resources. Such quantum enhancements have already been demonstrated in real-life devices, such as the advanced Laser Interferometer Gravitational-Wave Observatory (LIGO) experiments~\cite{Aasi2013}, where the sensitivity in the detection of gravitational waves has been enhanced through the use of squeezed light.

To understand the sensitivity limits that can be achieved through a quantum enhancement, it is necessary to know the minimum uncertainty in the estimation of the parameter of interest as it provides a measure of the minimum change in the parameter that can be detected. For a given state and system under study, the minimum variance of the mean of the parameter to estimate is bounded from below by the quantum Cram\'er-Rao bound (QCRB)~\cite{Helstrom1976,Holevo1982,PARIS2009,Tan2019}. Despite the quantum descriptor of this bound, it is not limited to quantum states but is derived from the quantum representation of the classical or quantum state used to probe the system under study and the use of quantum techniques to optimize over all possible measurements. Thus, the QCRB is independent of the measurement performed on the probing state after interacting with the system and depends only on the response of the system to the parameter of interest and the state probing the system. As a result, if a given measurement uncertainty saturates the QCRB, then that measurement is the optimal one and no other measurement can provide a further reduction in uncertainty. Furthermore, the ratio between the QCRBs for the probing quantum state and the optimal classical state establishes the maximum quantum enhancement that can be achieved. Therefore, measuring a parameter at the QCRB with a quantum state ensures the maximum sensitivity and quantum enhancement for that state.

Here we focus on the estimation of transmission, which is the basis of many sensing applications that have benefitted from the use of quantum states.  For example, quantum states of light have enhanced plasmonic sensors~\cite{Dowran2018}, two-photon absorption spectroscopy~\cite{Silberberg2009, Yao2012,Zheltikov2020}, and the calibration of the quantum efficiency of detectors~\cite{Penin1991, Migdall1995, Brida2006, Brida2006a, Worsley2009, AGAFONOV2011, Marino2011, Perina2012, Haderka2014}. For transmission estimation, it has been shown theoretically that single-mode states with reduced intensity noise~\cite{Monras2007,Adesso2009} and two-mode states with reduced intensity-difference noise~\cite{Invernizzi2011,Woodworth2020} can provide a quantum-based enhancement. Specifically, it is known that the Fock state~\cite{Adesso2009} and vacuum two-mode squeezed state (vTMSS)~\cite{Invernizzi2011} have the lowest possible QCRB for transmission estimation, referred to here as the ultimate bound. However, these states can only be generated at very low power levels~\cite{Agafonov2010, Chekhova2015, Loredo2016, Wang2016, Uria2020}. Since the ultimate QCRB scales inversely with the number of photons~\cite{Adesso2009, Invernizzi2011}, the low photon numbers of these states limits the absolute uncertainty in transmission estimation that can be achieved. As a result, these states are, in general, not able to surpass the corresponding classical state-of-the-art and their applicability to real-life sensing applications is limited. To overcome this limitation, it is possible to use bright quantum states of light that can be generated with orders magnitude larger number of photons. While such states are not able to reach the ultimate bound in general, they can achieve  a lower overall QCRB and surpass the classical state-of-the-art~\cite{Dowran2018}. Here we specialize to the use of the bright two-mode squeezed state (bTMSS), as it approaches the ultimate bound at high levels of squeezing~\cite{Woodworth2020} and can be generated at high powers~\cite{McCormick2008, Marino2008, Guerrero2020}. Thus, in practice, the bTMSS gives a better absolute estimation of transmission than the Fock or vTMSS given that it is a macroscopic quantum state. In addition, we have previously identified a measurement that can saturate the QCRB and can be implemented with current technology~\cite{Woodworth2020}.

The importance of quantum enhanced transmission estimation has led to recent experimental works that demonstrated that a quantum advantage is possible~\cite{Moreau2017, Whittaker2017, Atkinson2021, Li2021}. In 2017, Moreau \emph{et.~al}~\cite{Moreau2017} showed that the vTMSS does lead to reduced uncertainty in transmission estimation compared to a coherent state, but their measurement was only able to achieve this for transmissions above 50\% and did not saturate the QCRB. In the same year, Whittaker \emph{et.~al}~\cite{Whittaker2017} were able to saturate the ultimate bound using a single photon source, though only over a limited transmission range between 10\% and 30\% where the quantum advantage is small. In 2021, Li \emph{et.~al}~\cite{Li2021} showed a quantum advantage with bTMSS for transmissions above 40\%. While an improvement over Moreau \emph{et.~al}, their measurements were not able to saturate the QCRB. Subsequently, Atkinson \emph{et.~al}~\cite{Atkinson2021} used a bright single-mode squeezed state to measure a transmission modulation peak height. Their measurement, nevertheless, was assumed to saturate the QCRB based on a theoretical analysis without an experimental proof. Their work focused on a single modulation depth and studied the degree of quantum advantage as a function of different squeezing levels and detection bandwidth. Our current work shows for the first time that it is possible to perform measurements that saturate the QCRB over a broad range of transmission levels, without any free parameters, with bright quantum states of light. The bright nature of the quantum states used results in absolute sensitivities many orders of magnitudes larger than the only previous experiment that operated at the QCRB only for low transmissions~\cite{Li2021}.

\section{Experimental Setup and Procedure}

\begin{figure*}[htb]
\centering
\includegraphics{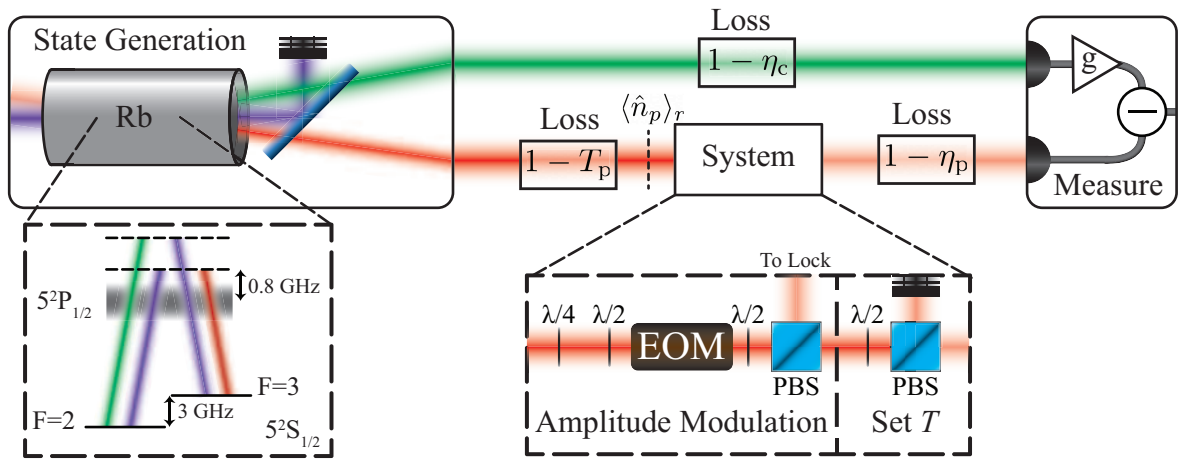}
\caption{Experimental setup for transmission estimation at the QCRB with a bTMSS. The bTMSS is generated in a $^{85}$Rb vapor cell via a FWM process in a double $\Lambda$ configuration in the D1 line, as shown in the ``state generation'' inset. A strong pump beam (shown in purple) is combined with a weak probe beam (shown in red) to generated the quantum correlated probe and conjugate (shown in green). The probe beam is used to probe the system under study, while the conjugate beam serves as the reference for the transmission estimation. We consider losses in the probe mode both before and after the system under study and losses in the conjugate mode. The ``system'' inset shows the configuration that is used to emulate a transmissive system. An electro-optic modulator (EOM) is used in an amplitude modulation configuration with the reflection from the polarizing beam splitter (PBS) after the EOM used to stabilize it, see Appendix~\ref{App:Locking}. After the amplitude modulation section of the system, a half waveplate and PBS are used to control the mean transmission through the system. The transmission of the system is given by the mean transmission of the EOM setup and the Set $T$ PBS. An optimal intensity-difference measurement of the probe and conjugate modes, with electronic attenuation of the photocurrent of the detected conjugate mode, is used to obtain the uncertainty in the estimation of the transmission, $\var{T}$.}
\label{fig:Setup}
\end{figure*}

For transmission estimation with a bTMSS, one mode is used to probe the system under study, and is therefore called the probe mode, while the other mode is used as a reference. For this study, we consider the number of photons in the probe mode that interact with the system, $\expected{\hat n_p}_r$ in Fig.~\ref{fig:Setup}, as the resources for the parameter estimation. Many systems have a limit in the number of photons they can interact with without damage or other adverse effects and this is the typical limiting factor for parameter estimation. It is for such systems that quantum states, which have a lower QCRB than classical states for the same number of probing photons, can provide a sensitivity enhancement that can surpass the classical state-of-the-art for practical applications.

The configuration that we use is shown in Fig.~\ref{fig:Setup}. We generate a bTMSS with a four-wave mixing (FWM) process, which is based on a double $\Lambda$ configuration, as shown in the ``state generation'' inset in Fig.~\ref{fig:Setup}. In this non-linear process, two photons from a strong pump beam are absorbed to simultaneously create one photon in the probe and one in a new beam commonly referred to as the conjugate, which serves as the reference for estimating the transmission. If the probe and conjugate modes are not seeded (input vacuum sates) then a vTMSS is generated. However, if either mode is seeded, typically with a coherent state, the generation rate of photons is increased.  If the power of the seeding mode(s) is large enough that the generation rate of stimulated photons is much larger than the rate for spontaneously generated photons, then the state is a bTMSS.

We implement the FWM process in the D1 line of $^{85}$Rb in a 12~mm long hot vapor cell heated to 120~$^\circ$C. A strong pump (600~mW of power and $1/e^2$ radius waist of 700~$\mu$m) is combined with the seeding probe mode (7~$\mu$W of power and $1/e^2$ radius waist of 400~$\mu$m) at an angle of 0.4$^\circ$ at the center of the Rb cell. The pump beam is generated with a Ti:Sapph laser at 795~nm while the seeding beam is generated by taking a portion of the pump and downshifting its frequency by 3.04~GHz via double passing an acousto-optic modulator (AOM). Before seeding the FWM process, a cleanup cavity (Newport SuperCavity model SR-140-C) is used to filter out any technical noise in the probe mode, such that it is shot noise limited at 1.5~MHz. For these parameters, the FWM has a gain of 11.4 and the generated probe and conjugate have a measured intensity-difference noise 8.0~dB below the shot noise, after subtracting the electronic noise. To keep the number of photons probing the system, $\expected{\hat n_p}_r$, constant throughout the experiment, we lock the probe seed power before the Rb vapor cell and stabilize the gain of the FWM by locking the temperature of the cell, the pump power, and the frequency of the laser. The frequency of the laser, and therefore of the pump and probe, is locked via the conjugate power, as explained in Appendix~\ref{App:Locking}.

To emulate a system with linear transmission, we use the configuration shown in the ``system'' inset of Fig.~\ref{fig:Setup}, which consists of two parts. The first part modulates the transmission, as needed to estimate the minimum resolvable change in transmission and thus the uncertainty in transmission estimation, and the second one sets the mean transmission through the system, $T$. To modulate the transmission we use an electro-optic modulator (EOM) in an amplitude modulation configuration. For light incident on the EOM with a polarization that is not aligned to one of its axis, the EOM introduces a phase shift between the field components in the directions of the EOM crystal axes.  This leads to a change in the polarization of the light that can be controlled by a voltage applied across the EOM crystal. When the EOM is followed by a half waveplate and a polarizing beam splitter (PBS), the polarization modulation is converted into a transmission modulation. A quarter waveplate and a half waveplate before the EOM give complete control over the polarization of the incident light, thus allowing for control of the transmission modulation properties. For the second part of the system, we use another half waveplate and PBS to explore the QCRB dependence on transmission. 

Finally, an optimized intensity-difference measurement is performed on the optical state, as we have previously shown theoretically that this measure saturates the QCRB for transmission estimation with a bTMSS~\cite{Woodworth2020}. This measurement is similar to a balanced intensity-difference measurement, where the measured photocurrents of the two modes are subtracted, except for an electronic attenuation of the photocurrent of one of the modes being performed before the subtraction. In our experiment, the photocurrent of the detected conjugate mode is attenuated to maximize the cancellation of the intensity noise of the detected probe mode.

\begin{figure}[htb]
\centering
\includegraphics{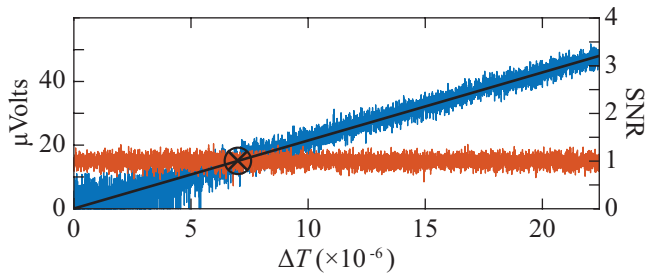}
\caption{Measured signal and noise for $T$=15\%, as a function of transmission modulation. The 
signal trace (blue), obtained while ramping the modulation,  and the optimized intensity-difference noise trace (red), obtained with the modulation off, measured with the spectrum analyzer are shown in volts on the left $y$-axis after subtraction of the electronic noise. The right $y$-axis shows the corresponding SNR obtained by taking the ratio of the signal to the mean value of the optimized intensity-difference noise. The value on the $x$-axis at which the signal is equal to the noise (or SNR=1), marked by a circled ``X'', gives the standard deviation in the estimation of transmission, $\Delta T$, for the bTMSS.}
\label{fig:Traces}
\end{figure}

To measure the minimum resolvable change in transmission, given by the standard deviation in transmission estimation ($\Delta T$), we use a spectrum analyzer to determine the point at which a calibrated transmission modulation (see Appendix~\ref{App:Calibrations}) is equal to the noise power of the optimized intensity-difference measurement in the absence of the modulation. At that point, the variance of the transmission modulation gives the uncertainty in transmission estimation. We find this point by ramping down the amplitude of the transmission modulation signal introduced with the EOM to determine the modulation at which the signal is equal to the noise, or the signal-to-noise-ratio (SNR) is equal to one, as indicated by the circled ``X'' mark in Fig.~\ref{fig:Traces}. We start with a large fixed modulation amplitude for 2~seconds, not shown in Fig.~\ref{fig:Traces}, to reduce the effects of ringing when the ramping cycle is repeated, followed by a linear reduction of the modulation amplitude to zero over 14~seconds. To obtain repeatable measurements, the EOM transmission modulation amplitude is locked (as described in Appendix~\ref{App:Locking}) during the entire procedure and the ramping is done by changing the setpoint of the lock, such that the EOM modulation voltage follows to produce a linear ramp of the transmission modulation amplitude. We measure the output of the optimized intensity-difference measurement with a spectrum analyzer in volts with and without the ramp. The signal, shown in blue in Fig.~\ref{fig:Traces} (volts scale on the left), is obtain by taking the measured trace with the ramp on and subtracting the mean optimized intensity-difference noise, which is given by the measured trace without the ramp (red trace in Fig.~\ref{fig:Traces}). The SNR, given on the right scale in Fig.~\ref{fig:Traces}, is obtained by the ratio in volts of the signal to the mean optimized intensity-difference noise. The obtained SNR is then fitted to a line (black solid line in  Fig.~\ref{fig:Traces}) to find the $\Delta T$ at which the SNR$=1$. This process is repeated at different transmissions $T$, set with the second part of the ``system'' defined in Fig.~\ref{fig:Setup}, to determine $\Delta T$ across a wide transmission range. During this process, the transmission is modulated with the EOM at 1.5~MHz given that at this frequency the probe seed beam for the FWM is shot noise limited after passing through the cleanup cavity.  This ensures that the measurements are not contaminated with technical noise and are thus dominated by the quantum statistics of the probing light, as needed to perform measurements at the QCRB.

We take a total of 20 sets of transmission uncertainty measurements. For each set we start at the maximum possible mean transmission, $\approx85\%$, and lower it in steps of $5\%$ to a minimum mean transmission of $10\%$.  At each transmission level we take one trace with the transmission modulation ramp on and one with it off in order to calculate the SNR.  We take one complete set of 16 mean transmissions and then return to the maximum transmission to take the next set. This approach allows us to rule out systematic effects, such as changes in the level of squeezing or probing power, that could also lead to changes in the measured uncertainties. Additionally, each mean transmission is measured for every set by first measuring the intensity of the probe mode before the system under study, thus reducing any biasing of the transmission due to power drifts.

As has been previously shown, the QCRB  for transmission estimation scales inversely with the number of probing photons~\cite{Adesso2009, Invernizzi2011,Woodworth2020}. Thus, in order to perform a direct comparison between the measured transmission uncertainties and the QCRB without any free parameters, a proper calibration of the number of photons used to probe the system under study is essential. We perform this calibration by measuring the photon flux (which is proportional to the probe optical power, set to 80~$\mu$W in the experiment) and multiplying it by the effective measurement time,$t$, for our setup, which is determined by the resolution bandwidth (RBW) of the spectrum analyzer. As outlined in Appendix~\ref{App:Count}, the effective measurement time for our spectrum analyzer is $\approx0.44$/RBW, which leads to $t=8.63~\mu$s for a RBW of 51~kHz used in the experiments.  This gives a mean photon number $\expected{\hat n_p}_r \sim 10^9$~photons.

\section{Results}

\begin{figure}[htb]
\centering
\includegraphics{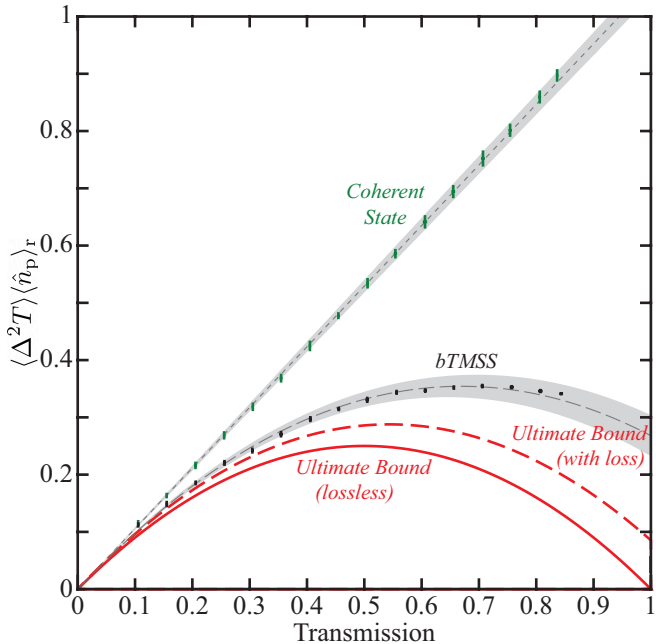}
\caption{Measured uncertainties in transmission estimation at the QCRB as a function of mean transmission $T$. The results are plotted as the product $\var{T}\expected{\hat n_p}_r$ to make the traces independent of the number of probing photons or resources used for the estimation. Black and green data points correspond to the bTMSS and coherent state measurements, respectively. The vertical lines, not always visible, around each data point are the one sigma standard deviation over the 20 measurements performed at each mean transmission. The dashed and dotted grey lines are the QCRB predictions for the bTMSS and coherent state, respectively, with the shaded grey regions giving the one sigma uncertainty in the predictions. No free parameters or fittings were used in the plots, which take into account all the experimental imperfections that were independently calibrated. The red solid and dashed lines denote, respectively, the ultimate bounds without and with probe losses.}
\label{fig:Results}
\end{figure}

The results for the bTMSS are shown as black data points (black dots with one sigma error bars) in Fig.~\ref{fig:Results}. To compare our measurement results with the ultimate bound, we consider, as a function of the mean transmission $T$, the product of the transmission estimation variance and mean number of probing photons, $\var{T}\expected{\hat n_p}_r$, which is independent of the number of photons (resources) used to probe the system. As an additional check to our procedure and to obtain a measure of the degree of quantum enhancement possible with the bTMSS, we repeat the experiment with the optimal classical configuration using a coherent state. To do so, we remove the Rb vapor cell but keep everything else the same between the measurements with a coherent state and a bTMSS. Since there is only one mode for the coherent state, the optimized intensity-difference measurement simplifies to an intensity measurement of the coherent state. The intensity measurement has also been shown to saturate the QCRB~\cite{Woodworth2020} for transmission estimation with a coherent state, thus we are comparing our measurements with the bTMSS to the best possible classical transmission estimation.  The results obtained with the optimal classical configuration are shown as green data points (green dots with one sigma error bars) in Fig.~\ref{fig:Results}.  As can be seen, for our maximum transmission of 84\% we obtained a quantum advantage over the optimal classical configuration by a factor of 2.6 through the use of a bTMSS with 8.0~dB of balanced intensity-difference squeezing.  This corresponds to a reduction of 62\% in the variance in the estimation of transmission. The error bars on the data points for both the bTMSS and the coherent state correspond to a one sigma standard deviation over the 20 measurements performed at each mean transmission.

To compare the measured transmission uncertainties with the QCRB, one needs to properly take into account the quantum state that is used to probe the system under study.  We previously showed that the QCRB for transmission estimation using a pure bTMSS, followed by losses in the probe mode before and after the system and loss in the conjugate mode (as shown in Fig.~\ref{fig:Setup}), is given by~\cite{Woodworth2020}
\be
\var{T}_\text{bTMSS}\ge\frac{T}{\eta_p\expected{\hat n_p}_r}-\frac{T^2}{\expected{\hat n_p}_r}T_pH_c\left[1-\sech(2s)\right],\label{eqn:bTMSS_QCRB}
\ee
where $\expected{\hat n_p}_r$ is the number of photons in the probe mode incident on the system, $T_p$ and $\eta_p$ are the transmissions before and after the system, respectively, and $s$ is the squeezing parameter that controls the rate of generation of photon pairs, i.e. the FWM gain, which also sets the quantum correlations between the probe and conjugate modes~\cite{Drummond2003, McCormick2006, McCormick2008}. Additionally
\be
H_c=\frac{\left(2\eta_c-1\right)\left[1+2\sinh^2(s)\right]}{1+2\eta_c\sinh^2(s)},
\ee
where $\eta_c$ is the transmission of the conjugate mode. It should nevertheless be pointed out that the assumptions leading to Eq.~\eqref{eqn:bTMSS_QCRB} are not exactly valid for our experimental implementation, as the generated state is not a pure bTMSS.  This is due to internal losses in the atomic medium used to implement the FWM and the fact that the two-mode squeezing operator does not commute with the loss operator, which means that we cannot consider the source as a perfect squeezer followed by the losses introduced by the atomic system.

To have a more accurate characterization of the generated quantum state, and thus correctly set the QCRB, we consider a model for the source that consists of an infinite series of alternating infinitesimal layers of two-mode squeezers and beam splitters (to model internal loss)~\cite{McCormick2008,Jasperse2011}. This allows us to obtain the covariance matrix, $\sigma$, and displacement vector, $\vec d$, needed to calculate the QCRB following the method for Gaussian states given by \v{S}afr\'{a}nek~\cite{Safranek2015}, where the uncertainty in the estimation of transmission satisfies
\be
\var{T}\ge\left(2\frac{\partial \vec d^\dagger}{\partial T}\sigma^{-1}\frac{\partial \vec d}{\partial T}\right)^{-1}
\ee
in the bright limit when the stimulated photon pair generation dominates over the spontaneous pair generation~\cite{Woodworth2020}. We assume there is no absorption of the conjugate mode due to the atomic medium as its frequency is far-off resonance from any transition. In this case, the QCRB for our quantum state is given by
\be
\var{T}_\text{bTMSS}\ge\frac{T}{\eta_p\expected{\hat n_p}_r}-\frac{T^2}{\expected{\hat n_p}_r}T_pH'_c\frac{32s^2\sqrt{T_\text{a}}\sinh^2\left(\frac{\xi}{4}\right)}{\xi^2(\sqrt{T_\text{a}}-1)+\Gamma},\label{eqn:bTMSS_QCRB_New}
\ee
where $T_\text{a}$ is the product of all the transmissions of the beam splitters in the probe mode used to the model the source, $s$ is the sum of all squeezing parameters of the squeezers in the model, $\xi=\sqrt{16s^2+\ln^2(T_\text{a})}$,
\bea
\Gamma&=&\sqrt{T_\text{a}} \Bigg\{\cosh\left(\frac{\xi}{2}\right)\left[\xi^2+\ln^2(T_\text{a})\right]\notag\\
&&-\ln(T_\text{a})\left[\ln(T_\text{a})+2\xi\sinh\left(\frac{\xi}{2}\right)\right]\Bigg\},
\eea
and
\be
H'_c=\frac{2\eta_c-1}{\eta_c}\left(1+\frac{\xi^2(\eta_c-1)}{\xi^2(1+\eta_c(\sqrt{T_\text{a}}-2))+\eta_c\Gamma}\right).
\ee
The factor $H'_c$ plays the same role as $H_c$ in the model described in Eq.~\eqref{eqn:bTMSS_QCRB}, such that $H'_c=1$ when $\eta_c=1$, $H'_c=0$ when $\eta_c=1/2$, and $H'_c<0$ when $\eta_c<1/2$.  The QCRB for the coherent state can then be obtained by setting $s=0$ in either Eq.~\eqref{eqn:bTMSS_QCRB} or~\eqref{eqn:bTMSS_QCRB_New}, to give
\be
\var{T}_\text{coh}\ge\frac{T}{\eta_p\expected{\hat n_p}_r},\label{eqn:cs_QCRB}
\ee
which scales linearly with transmission.

To evaluate the QCRB for our system, we need to consider the losses external to the system that do not form part of the measured transmission uncertainties shown in Fig.~\ref{fig:Results},
but increase the QCRB, as can be seen from Eqs.~\eqref{eqn:bTMSS_QCRB} and \eqref{eqn:bTMSS_QCRB_New}. Imperfect probe transmission before the system, $T_\text{p}$, comes from the Rb vapor cell output window, polarization filter used to separate the pump mode from the probe and conjugate modes, and the various mirrors and lenses used to propagate the probe mode to the system under study. The transmission through the cell window was measured to be $98.8\%\pm1$\% and the propagation transmission between the system and Rb vapor cell, after the cell window, was measured to be $98.4\%\pm1$\% for a total transmission before the system, $T_\text{p}$, of $97.3\%\pm1\%$. Transmission in the path of the probe after the system, $\eta_p$, comes mainly from the photodiode's quantum efficiency, which we approximate from the data sheet and previous measurements to be of $94.5\%\pm2$\%~\cite{McCormick2008}. The conjugate mode transmission, $\eta_c$, is equal to the combination of the probe transmission both before and after the system under study, $\eta_c=T_p\eta_p$, as the probe and conjugate modes share many optical elements and the quantum efficiencies of the photodiodes are the same for the probe and conjugate. This leads to a total conjugate mode transmission, $\eta_c$, of $91.9\%\pm2\%$. We also need to estimate the effective squeezing parameter $s$ and probe loss due to atomic absorption $T_{a}$.  We do so by measuring the balanced intensity-difference noise and the single beam intensity noises of the probe and conjugate modes by going around the system under study. We then compare these values, after backtracking the propagation and detection losses ($T_p$, $\eta_p$, and $\eta_c$), with the corresponding values obtained from the model of the source composed of layers of squeezers and losses to find the optimal parameters of the source (see Appendix~\ref{App:Infer} for the optimization procedure).  Following this procedure we find values of $s=2.04\pm0.02$ and $T_a=71\%\pm2\%$.

The dashed and dotted grey lines in Fig.~\ref{fig:Results} correspond to the QCRB predictions given by Eqs.~\eqref{eqn:bTMSS_QCRB_New} and~\eqref{eqn:cs_QCRB} for the bTMSS  and coherent state, respectively, after taking into account all the experimental imperfections. The shaded regions around these lines represent the theoretical one sigma uncertainty in the QCRB due to uncertainties in the estimation of the losses and the parameters of the generated bTMSS. As can be seen, the measured data is well within the predicted QCRB for both the bTMSS and coherent state, which shows that the measurements performed saturate the QCRB for transmission estimation over the accessible transmission range without any free parameters in the theory.  This indicates that the transmission estimation measurements performed are optimal and no further enhancements are possible with the optical states that are used.

One of the reasons we consider the use of a bTMSS is that it approaches the ultimate bound in transmission estimation as the level of squeezing increases, even for the imperfect bTMSS generated by our FWM source.  This can be seen if one takes the limit of infinite squeezing, $s\rightarrow\infty$, and perfect conjugate detection, in Eq.~\eqref{eqn:bTMSS_QCRB} or~\eqref{eqn:bTMSS_QCRB_New}.  In this limit the equations reduce to
\be
\var{T}_\text{ult}\ge\frac{T}{\eta_p\expected{\hat n_p}_r}-\frac{T^2}{\expected{\hat n_p}_r}T_p,\label{eqn:ult_QCRB}
\ee
which corresponds to the ultimate bound in transmission estimation~\cite{Woodworth2020}. The red lines in Fig.~\ref{fig:Results} correspond the ultimate bound, with the solid red line giving the ultimate bound for the lossless case and the dashed red line giving the ultimate bound in the case in which the losses on the probe mode before ($T_p$) and after ($\eta_p$) the system under study are the same as those in our experimental implementation. While we are not at the ultimate bound, we can see that the bTMSS does approach it. For example, at our maximum transmission of 84\%, where we obtain the maximum quantum enhancement in transmission estimation, we are only a factor of around 1.7 away from the ultimate bound. Furthermore, if we were to compare the use of a Fock state, for which the ultimate bound can be achieved, to probe the system under study, our absolute sensitivity would be $\sim 10^{9}$ times larger due to the large number of probing photons ($\expected{\hat n_p}_r$) in a bTMSS.

\section{Summary}
We performed transmission estimation measurements at the QCRB for both a bTMSS and a coherent state with simple measurement techniques. We were able to saturate this fundamental limit across a broad transmission range of 85\% to 10\%. While the bTMSS we generated was a factor of 1.7 away from the ultimate bound given by Fock states or vTMSS, we were still able to show a 62\% reduction in the variance in transmission estimation with respect to the optimal classical configuration at 84\% transmission when using a bTMSS with 8.0~dB of balanced intensity-difference squeezing.  Furthermore, the large number of photons with which a bTMSS can be generated leads to an absolute sensitivity in transmission estimation many orders of magnitude larger that the one that can be achieved with either a Fock state or a vTMSS. Given the applicability of transmission estimation to a number of sensing protocols and that the required measurements to saturate the QCRB are readily available, the results presented here are expected to enable quantum-enhanced sensors that can surpass the classical state-of-the-art, and promise to have significant impact to a number of fields.

\section*{Acknowledgments}

This work was supported by the W. M. Keck Foundation and by the National Science Foundation (NSF) (Grant No. PHYS-1752938). C.H-A. acknowledges support from Fondecyt Grant Nº 11190078, and Conicyt-PAI grant 77180003, and ANID - Millennium Science Initiative Program - ICN17-012. The authors would also like to thank Kellen Lawson for useful discussions on multi-parameter fitting and uncertainties.

\appendix

\section{Locking}
\label{App:Locking}

To preform measurements that are able to saturate the QCRB, we need to stabilize multiple aspects of the experiment such as the number of photons used to probe the system under study and the transmission modulation amplitude.  To keep the number of probing photons constant, the power of the probe beam after the FWM process needs to be stabilized.  This requires keeping both the seed probe power and the gain of the FWM processes stable.

To stabilize the power of the seed probe for the FWM process, a portion of the probe beam is picked off via a half waveplate and PBS before the Rb vapor cell. This pick off is then detected with a photodiode and serves as the error signal.  The power of the seed probe is kept constant by controlling the diffraction efficiency of the AOM used to generate the probe from the pump beam. This setup is typically referred to as a noise-eater.

Given that the gain of the FWM process depends on the atomic number density, pump power, and frequencies of the involved fields, all of these have to be stabilized.  The atomic number density depends on the temperature of the Rb vapor cell. Thus, a temperature controller is used to stabilize the Rb cell temperature to 120~$^\circ$C within a fraction of a degree. As with the probe beam, the pump power is locked before the Rb vapor cell by detecting a small portion of the beam that is picked off via a beam sampler. The pump power is then kept constant by feeding back to an electronically controlled rotation mount containing a half waveplate placed before a PBS. This feedback control is slow, however the pump power changes are mostly due to slower thermal drifts and the intensity noise of the pump has little effect on the generated quantum state. Thus, a high bandwidth noise-eater like the one used for the probe, is not needed.  Finally, since a change in the frequency of the laser changes the gain and thus the output conjugate and probe power, we use the conjugate power as an error signal to compensate for the frequency drifts.  We use this approach as the probe transmission, and thus detected power, is changed as part of the experiment. The error signal from the conjugate power is then fed back to lock the laser frequency, which results in a stable lock over the more than 20 hours needed to take the data.

\begin{figure}[htb]
\centering
\includegraphics{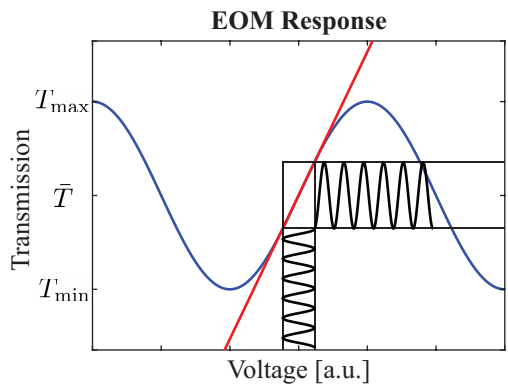}
\caption{EOM setup response. The transmission through the EOM setup as a function of applied voltage across the EOM crystal is shown in blue. The red line marks the linear region of the EOM around the operating voltage used in the experiment, which is centered around $\bar T$. A modulation of the voltage around this point, shown in black, results in a transmission modulation at the same frequency, as long as the modulation is within the linear regime.}
\label{fig:VppToDT}
\end{figure}

The transmission modulation amplitude must also be controlled for reproducibility from data set to data set. This requires tight control of the transmission modulation amplitude ramp and control over the mean transmission through the EOM. In the implementation of the system, as shown in the ``system'' inset of Fig.~\ref{fig:Setup}, the EOM setup has a quarter waveplate and a half waveplate before the EOM and a half waveplate and a PBS after the EOM. The waveplates before the EOM, used for the amplitude modulation configuration, are set such that the EOM has a high transmission and the transmission modulation of the system ($\Delta T$) is within the linear regime, see Fig.~\ref{fig:VppToDT}. As such, the maximum ($T_\text{max}$) and minimum ($T_\text{min}$) transmissions through the EOM setup have to be properly set. For our experiment the ratio $T_\text{max}/T_\text{min}$ is set to around 1.02 with a mean transmission of 84\%, giving a maximum change in transmission through the EOM of $\approx0.8\%$ via an applied voltage. To lock the EOM, we detect the output reflection of the PBS of the amplitude modulation configuration. The output of the photodetector is split into DC and AC signals to lock the mean transmission and transmission modulation amplitude, respectively. The DC lock ensures the EOM operates in the linear regime by keeping the mean transmission at its center value, see Fig.~\ref{fig:VppToDT}. Locking the mean transmission at the center point, $\bar{T}$ in Fig.~\ref{fig:VppToDT}, provides the largest possible slope to maximize the transmission modulation amplitude for a given modulation voltage. Furthermore, if the mean transmission drifts, the calibration of the transmission modulation amplitude (see Appendix~\ref{App:Calibrations}) is no longer valid. The AC lock is used to control the transmission modulation amplitude during the amplitude ramp. The AC lock PID output is sent to the function generator used to create the modulation voltage sent to the EOM. The AC and DC voltage outputs from the locking electronics are combined with a Bias Tee. The combined signal is then amplified and sent to the EOM.

Another important aspect of being able to reach the QCRB is to filter out the classical technical noise in the optical state used to probe the system under study. While the power of the seed probe beam was stabilized with a noise eater, such a configuration is unable to reduce the intensity noise to the shot noise limit~\cite{Bachor2004}.  As a result, a cleanup cavity (Newport SuperCavity model SR-140-C) is used to reach the shot noise limit at our operating frequency of 1.5~MHz. We use the Pound-Drever-Hall~\cite{Drever1983,Black2001} locking technique to keep the cleanup cavity on resonance with the probe. An EOM is used to add a phase modulation to the probe mode before coupling it into the cleanup cavity.  The modulation frequency is set to 10~MHz, which is significantly larger than the linewidth of the cavity ($<0.6$~MHz).  As a result, the sidebands from the modulation are not transmitted by the cavity and the reflected light can be used to generate the required error signal.

\section{Calibrations}
\label{App:Calibrations}

In order to compare the measured transmission uncertainties with the theoretical QCRB without any free parameters, the photon flux, propagation transmissions, and transmission modulation amplitude need to be properly calibrated.

To measure the propagation transmission we use two power meters. The first power meter, PM-A, is set on a flip mount in front of the Rb cell while the other, PM-B, is used to measure the transmission. First, we perform a relative calibration of the two power meters. We do this by placing PM-B right behind PM-A and measuring the power multiple times with both power meters by flipping PM-A in and out of the beam path. The ratio of the measured powers are then used to remove any systematic bias in the measurements. We then measure the transmission of the probe mode before the system under study without the Rb cell in place to be $T_\text{common}=98.4\%\pm1\%$. This path has many common optical elements for the probe and conjugate mode paths. We then place the Rb cell back in place and, working off resonance, find the transmission for each of the cell windows to be $T_\text{window}=98.8\%\pm1\%$. The quantum efficiency of the photodiodes are taken from previous calibration results to be $\eta=94.5\%\pm2\%$~\cite{McCormick2008}. Altogether this leads to a probe transmission before the system of $T_p=T_\text{common}T_\text{window}=97.3\%\pm1\%$, probe transmission after the system of  $\eta_p=94.5\%\pm2\%$, and a conjugate transmission of $\eta_c=T_\text{common}T_\text{window}\eta=91.9\%\pm2\%$.

\begin{figure}[htb]
\centering
\includegraphics{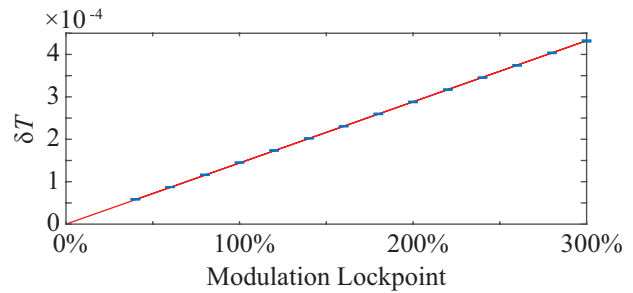}
\caption{Calibration of the change in transmission through the EOM setup, $\delta T$, as a function of the modulation lockpoint of the EOM. The calibration data is shown in blue with one sigma error bars for the $y$-axis as the modulation lockpoint is scanned from 0\% to 300\%, with the scale set by the range of lockpoints used for the experiment. The red line gives the linear fit to the data and shows that even at three times the modulation depth used in the experiment the EOM response remains linear.}
\label{fig:DeltaTCal}
\end{figure}

We operate the EOM in a regime in which the transmission modulation it introduces ($\delta T$) is linear with the voltage applied across the EOM crystal. To calibrate the slope and thus the relation between $\delta T$ and the applied voltage, we record the probe power oscillations on an oscilloscope for a given set of 14 modulation amplitude lockpoints. We then perform a fast Fourier transform of the recorded traces to isolate the amplitude of the oscillation from the noise and fit the transmission modulation to the applied voltage oscillation amplitude, see Fig.~\ref{fig:DeltaTCal}. We consider a maximum modulation lockpoint of 300\%, corresponding to a value three times higher than the maximum lockpoint value used in the experiment, due to the high electronic noise of the oscilloscope as compared to the spectrum analyzer. As can be seen in Fig.~\ref{fig:DeltaTCal}, the transmission modulation amplitude is linear with the applied voltage even at the higher lockpoints used for the calibration.

Finally, the transmission modulation due to the whole system under study ($\Delta T$) for a given mean transmission, $\bar T$, is given by $\Delta T=\bar T\delta T$. As the mean transmission is reduced during each data set, the transmission modulation also decreased. This leads to the difference between the $x$-axis of Fig.~\ref{fig:Traces} and the $y$-axis of Fig.~\ref{fig:DeltaTCal} in the range between 0\% and 100\% of the modulation lockpoint.

\section{Photon Counting}
\label{App:Count}

The dependence of the QCRB on the number of photons used to probe the systems under study makes it necessary to estimate such a quantity for the bTMSS, which is a continuous state. To go from a continuous photon flux to a discrete photon number requires bucketing the flux into discrete measurement time bins. The effective measurement time, $t$, for such a time bin is given by the response time of the measurement apparatus. In our experiments the measurement response is dominated by spectrum analyzer, such that the effective measurement time is set by its RBW. To find the effective measurement time for a given RBW, we first find the relationship between the variance measured with the spectrum analyzer and the number variance of the detected optical state. We then take advantage of the known relationship between the number variance and mean photon number for the coherent state, that is $\var{\hat n}=\expected{\hat n}$.

\begin{figure}[htb]
\centering
\includegraphics{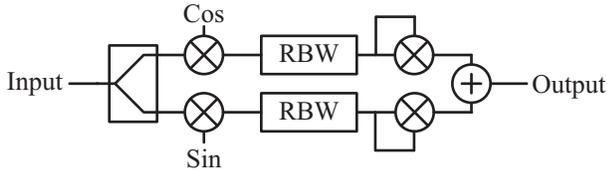}
\caption{Schematic of a basic spectrum analyzer. For a given input $I$, the spectrum analyzer splits the signal into two channels and mix each of resulting signals with an electronic local oscillator (LO). The LOs used for the two channels are 90$^\circ$ out of phase to obtain the in phase and out of phase components. After mixing with the LO, the resulting signal in each channel is passed through a low pass filter, which represents the RBW filter of the spectrum analyzer, and then squared. Finally, the signals from both channels are summed to obtain the output. An additional filter, the video bandwidth filter, can be used on the output to reduce the fluctuations on the noise measurements.  This filter, however, does not affect the mean value of the measured noise, and as a result it is not considered given that it does not play a role in our results.}
\label{fig:SA}
\end{figure}

We consider the basic spectrum analyzer design shown in Fig.~\ref{fig:SA} to determine $t$. Since the output, $O$, of the spectrum analyzer is proportional to the variance of the input, $I$, we first determine the proportionality constant, $K$, between these two quantities. In the experiment, the value of this constant has no effect on the number of photons measured, as it is purely a scaling factor due to the electronics and therefore needs to be taken into account.  To find the proportionality constant, we first consider a deterministic signal $I_\text{dtm}=A\sin(2\pi ft+\phi)$ with variance $\var{I_\text{dtm}}=A^2/2$. If we set the frequency of the electronic local oscillator (LO) used to demodulate the in phase ($\cos$) and out of phase ($\sin$) components of the input to the same frequency as our deterministic signal,  we get - after splitting, LO mixing, resolution bandwidth filtering, squaring, and summing (see Fig.~\ref{fig:SA}) - an output of the form
\be
\expected{O_\text{dtm}}=\frac{A^2}{8}\left|H(0)\right|^2=K\frac{A^2}{2},
\label{eqn:DetPeak}
\ee
where $H(f)$ is the frequency response of the RBW filter. As a result,
\be
K=\frac{1}{4}\left|H(0)\right|^2,
\ee
such that the ratio $\expected{O}/K$ gives the actual variance of the input, $\var{I}$. The number variance for an input state can then be related to the output of the spectrum analyzer through error propagation according to
\be
\var{\hat n}=\var{I}\bigg{/}\left|\frac{\partial \expected{I}}{\partial \expected{\hat n}}\right|^{2}
=\frac{\expected{O}}{K}\bigg{/}\left|\frac{\partial \expected{I}}{\partial \expected{\hat n}}\right|^{2}.\label{eqn:NumVar}
\ee

Next, we consider an input state given by a coherent state to take advantage of the relation between the number variance and the mean number of photons to calculate the effective integration time $t$.  To do so we need to calculate the different terms on the right-hand-side of Eq.~\eqref{eqn:NumVar}. The mean value for an input coherent state is given by
\be
\expected{I_\text{coh}}=C_{p\rightarrow i}|\alpha|^2=C_{p\rightarrow i}\frac{\expected{\hat n}}{t},\label{eqn:CohInt}
\ee
where $C_{p\rightarrow i}$ is the gain of the photodetector, $|\alpha|^2$ is the mean photon flux of the coherent state, and $\expected{\hat n}$ is the average number of photons detected over measurement time $t$. To find the mean value of the output of the spectrum analyzer, which in this case corresponds to noise power of the coherent state, we can define the fluctuation operator $\delta \hat a=\hat a-\alpha$ such that $\delta \hat{I}_\text{coh}=C_{p\rightarrow i}\delta \hat a^\dagger\delta \hat a$. The expectation value of the output can then be shown to be given by
\be
\expected{O_\text{coh}}=\frac{|\alpha|^2C_{p\rightarrow i}^2}{2}\int_{-\infty}^\infty \left|H(f)\right|^2 df.
\label{eqn:CohNoise}
\ee
As a result, the variance of the input takes the form
\be
\var{I_\text{coh}}=\frac{\expected{O_\text{coh}}}{K}=2|\alpha|^2C_{p\rightarrow i}^2\frac{\int_{-\infty}^\infty \left|H(f)\right|^2 df}{\left|H(0)\right|^2}.\label{eqn:NormCohNoise}
\ee
We can see that Eq.~\eqref{eqn:CohNoise} is proportional to $\int_{-\infty}^\infty \left|H(f)\right|^2 df$ while the output from a deterministic modulation input, Eq.~\eqref{eqn:DetPeak}, is proportional to $\left|H(0)\right|^2$. This difference comes from the deterministic signal being a single frequency peak, such that the contribution to the power spectrum (output of spectrum analyzer) is dominated by the power in the deterministic signal. On the other hand, the intensity noise of the coherent state is broadband, such that the power is distributed over all frequency components of the system response.

To find the measurement time, $t$, we can use Eqs.~\eqref{eqn:CohInt} and~\eqref{eqn:NormCohNoise} in Eq.~\eqref{eqn:NumVar} to find the number variance for a coherent state
\bea
\var{\hat n}_\text{coh}&=&2\expected{\hat n}t\frac{\int_{-\infty}^\infty \left|H(f)\right|^2 df}{\left|H(0)\right|^2}\label{eqn:GetTime}\\
&=&\expected{\hat n},\label{eqn:CohNumVar}
\eea
where the last line is a property of the coherent state. We can now set the right-hand-side of Eq.~\eqref{eqn:GetTime} equal to Eq.~\eqref{eqn:CohNumVar} to show that the effective measurement time is given by
\be
t=\frac{\left|H(0)\right|^2}{2\int_{-\infty}^\infty \left|H(f)\right|^2 df}.
\ee

For an ideal spectrum analyzer, the RBW filter has a Gaussian profile with the full-width at half-maximum given by the value of the RBW, such that
\be
\frac{\left|H_\text{gaus}(f)\right|^2}{\left|H_\text{gaus}(0)\right|^2}=e^{-\frac{4\ln(2)f^2}{\text{RBW}^2}}.
\ee
Therefore, the time for our measurements would ideally be
\be
t_\text{gaus}=\sqrt{\frac{\ln(2)}{\pi}}\frac{1}{\text{RBW}}\approx\frac{0.47}{\text{RBW}}.
\ee
However, real spectrum analyzers are only able to approximate a Gaussian filter. For our spectrum analyzer (Agilent model E4445A) the filter is a 4-pole synchronously tuned filter with a correction factor of $\approx~0.94$~\cite{Keysight2017}. Thus, the actual effective measurement time for our system is given by $t\approx0.44/\text{RBW}$.

Finally, to obtain the number of photons used in the experiment, we need to calibrate the photon flux, $\Phi=\expected{\hat n}/t$, for the probe power used in the experiment. The DC voltage output, $V_\text{dc}$, of the photodetectors used to detect the probe and conjugate modes are linearly dependent on the optical power, $P$, detected, such that $V_\text{dc}=mP$ with $m$ giving the proportionality constant. We first calibrate $m$ by flipped a power meter in and out of the beam path in front of the photodetector and measuring the optical power $P$ and output voltage $V_\text{dc}$ for incident beams with different powers. This allows us to perform a linear fit to get an accurate measure of $m$. Thus, we can find the photon flux according to $P=\Phi\frac{hc}{\lambda}$, where $\lambda$ is the wavelength of the probe mode (795 nm for our experiment), $h$  is Planck's constant, and $c$ is the speed of light. The photon number is then given by
\be
\expected{\hat n}=\Phi t=\frac{\lambda}{hc}\frac{t}{m}V_\text{dc},
\ee
where $V_\text{dc}$ is recorded for each transmission for each data set measured to determine the mean transmission $T$ independently for each data point taken. The number of probing photons is calculated by bypassing the system under study and is also measured for each data point taken.

\section{Inferring the Squeezing Parameters}
\label{App:Infer}

The effective squeezing parameter, $s$, and the transmission of the probe through the Rb cell, $T_a$, must both be estimated to find the state generated. The probe transmission measured without the pump field is not an accurate estimation of $T_a$ as the strong pump field leads to optical pumping, which modifies the transmission of the probe. To estimate $s$ and $T_{a}$ we measure the noise properties of the generated bTMSS and compare the measured values with the theoretically calculated noise properties that take into account the distributed losses in the atomic medium.

For the generated bTMSS, we measure the balanced intensity-difference noise and the individual intensity noises of the probe and conjugate modes at 1.5 MHz, the operating frequency of the experiment, and subtract the electronic noise. These noises are then normalized by the corresponding shot noise and backtracked to obtain the normalized noises generated by the FWM process.  We backtrack the noises by removing the effects of the loss from the Rb cell output window, optical path to the detectors, and the quantum efficiency of the detectors. This is done by using the relation
\be
\text{N}_0=\frac{\text{N}_\text{m}-(1-\eta)}{\eta},
\ee
where N$_0$ is the normalized noise directly generated by the source, N$_\text{m}$ is the measured normalized noise, and $\eta$ is the total transmission, which is given by $T_\text{p}\eta_\text{p}$ for the probe beam and $\eta_\text{c}$ for the conjugate beam.  As described above, both transmission are the same so we can take $\eta=\eta_\text{c}=T_\text{p}\eta_\text{p}$ to backtrack the intensity-difference noise.

To calculate the theoretically expected noise properties that take into account distributed losses in the source, we model the FWM in the atomic system as an infinite series of infinitesimal layers of two-mode squeezers and beam splitters (to take into account losses)~\cite{McCormick2008, Jasperse2011}.  Given that the frequency of the conjugate beam is far away from any atomic resonance, we assume that it does not experience any losses.  Thus, $T_a$ represents the loss due to atomic absorption for the probe beam. The sum of all the infinitesimal squeezing parameters gives the effective value of $s$ and the product of all the infinitesimal transmissions of the beam splitters gives $T_a$. For the theoretical normalized noises, we use the results given in~\cite{Jasperse2011} and verify them through the numerical approach given in~\cite{McCormick2008}. The analytical solutions for the theoretical model for the normalized noises are given by
\bea
\frac{\var{\!\left(\hat n_p\!\!-\!\hat n_c\right)}}{\expected{\hat n_p}+\expected{\hat n_c}}&=&1-\frac{2s\sinh^2\left(\frac{\xi}{4}\right)}{\xi\cosh\left(\frac{\xi}{2}+\zeta\right)}\notag\\
&&-\sqrt{T_a}\frac{s\ln^2(T_a)\sinh^4\left(\frac{\xi}{4}\right)}{2\xi^3\cosh\left(\frac{\xi}{2}+\zeta\right)}
\eea
\be
\frac{\var{\hat n_p}}{\expected{\hat n_p}}=\frac{16s^2\left\{1-\sqrt{T_a}\left[1-\cos\left(\frac{\xi}{2}\right)\right]\right\}+\ln^2(T_a)}{\xi^2}
\ee
\bea
\frac{\var{\hat n_c}}{\expected{\hat n_c}}&=&\frac{16s^2\sqrt{T_a}}{\xi^2}-1\notag\\
&&\!\!\!\!\!\!\!\!\!\!\!\!\!\!\!\!-\frac{2\sqrt{T_a}\left[\left(8s^2-\xi^2\right)\cosh\left(\frac{\xi}{2}\right)+\xi\ln(T_a)\sinh\left(\frac{\xi}{2}\right)\right]}{\xi^2},\notag\\
\eea
where $\xi=\sqrt{16s^2+\ln^2(T_a)}$ and $\tanh(\zeta)=\ln(T_a)/\xi$.

We then determine $s$ and $T_a$ by finding the values of these parameters that provided the best match between the theoretical model and measurements, in log scale, through the goodness-of-fit parameter $\chi^2$~\cite{Bevington2003},
\be
\chi^2=\sum_i\frac{\left[\text{Measurement}_i-\text{Theory}_i(s,T_a)\right]^2}{\text{Variance of Measurement}_i},
\ee
where the sum is over the three normalized noises. We minimize $\chi^2$ using a differential evolution optimization algorithm~\cite{Storn1997, Neri2009} to find $s=2.04$ and $T_a=71\%$ with a $\chi^2=0.4563$. The uncertainty in the fitted parameters is found by varying their values until $\chi^2$ increases by the reduced $\chi^2$. The reduced $\chi^2$ is $\chi^2/\text{dof}$, where dof gives the degrees of freedom given by the number of measurements minus the number of parameters to fit. For our case the dof is 1, so $\chi^2/\text{dof}=\chi^2$. Thus, we determine the values of the parameters for which $\chi^2$ is doubled to find the uncertainties for the parameters. Altogether, we find $s=2.04\pm.02$ and $T_a=71\%\pm2\%$.

The optimization algorithm we used performs a differential evolution, which is a type of genetic algorithm. As such, it starts with a random set of possible solutions, tests how well each solution works using some goodness-of-fit parameter, and then mixes the solutions in an attempt to increase the goodness-of-fit. Here, minimizing $\chi^2$ serves as our goodness-of-fit. To initialize the algorithm, we randomly created a population with 5,000 points of $s$ and $T_a$ values limited to $0\le s\le3$ and $0.5\le T_a\le1$. We then record the $\chi^2$ value for each of these points.

After initializing the population, we mix the different elements in a way that optimizes towards the lowest $\chi^2$ value. To do this we first find the optimal point, $P_o$, that has values of $s$ and $T_a$ that give the lowest $\chi^2$ of all the points in the population. We take this point to make new points for the next generation of population. We select one of the remaining 4,999 points to possibly replace, $P_i$. To create the possible replacement point, $P_r$, we randomly select two more points from the current population, $P_j$ and $P_k$, such that $P_o\ne P_i\ne P_j\ne P_k$. We then create a vector pointing from $P_j$ to $P_k$ normalized by the limits placed on the parameters, that is, normalized by $\sqrt{(0-3)^2+(.5-1)^2}$ for our case. We create the replacement point $P_r$ by adding this vector to $P_o$. If the new point is outside of the limits set for $s$ and $T_{a}$ for one or both of the parameter values, the point is set to the closest limit. We then find $\chi^2$ for $P_r$ and, if it is lower than the $\chi^2$ of point $P_i$, we replace $P_i$ with $P_r$. Finally, we repeat this for all points that are not $P_o$, randomly picking new $P_j$ and $P_k$ for each one. To increase the algorithm's ability to find global minimums in the presence of local minimums, we perform the replacement for only 70\% of the time when $P_r$ is better than $P_i$.

There is no set limit to the number of iterations needed to find the minimum $\chi^2$, so we iterate the algorithm until the spread of the population is orders of magnitude less than the uncertainty in the parameter values. Saving the $\chi^2$ value for each point during the differential evolution gives a look at the $\chi^2$ dependence on $s$ and $T_a$, especially around the global minimum. Thus, we can find a circular region around the minimum $\chi^2$ where the value is doubled to find the uncertainty in the transmission and squeezing parameter.

\end{document}